\documentclass[12pt,preprint]{elsarticle}
\usepackage[latin9]{inputenc}
\usepackage{float}
\usepackage{amstext}
\usepackage{amssymb}
\usepackage{graphicx}

\makeatletter

\providecommand{\tabularnewline}{\\}

\usepackage{url}

\newcounter{bla}

\journal{Computer Physics Communications}

\makeatother

\begin{document}
\begin{frontmatter}



\title{ZMCintegral: a Package for Multi-Dimensional Monte Carlo Integration
on Multi-GPUs}


\author[a]{Hong-Zhong Wu\corref{author}}

\author[a]{Jun-Jie Zhang\corref{author}}

\author[b,c]{Long-Gang Pang}

\author[a]{Qun Wang}

\cortext[author]{Both authors contributed equally to this manuscript.}

\address[a]{Department of Modern Physics, University of Science and Technology
of China}

\address[b]{Physics Department, University of California, Berkeley, CA 94720,
USA and}

\address[c]{Nuclear Science Division, Lawrence Berkeley National Laboratory,
Berkeley, CA 94720, USA}


\begin{abstract}
We have developed a Python package ZMCintegral for multi-dimensional
Monte Carlo integration on multiple Graphics Processing Units(GPUs).
The package employs a stratified sampling and heuristic tree search
algorithm. We have built three versions of this package: one with
Tensorflow and other two with Numba, and both support general user
defined functions with a user-friendly interface. We have demonstrated
that Tensorflow and Numba help inexperienced scientific researchers
to parallelize their programs on multiple GPUs with little work. The
precision and speed of our package is compared with that of VEGAS
for two typical integrands, a 6-dimensional oscillating function and
a 9-dimensional Gaussian function. The results show that the speed
of ZMCintegral is comparable to that of the VEGAS with a given precision.
For heavy calculations, the algorithm can be scaled on distributed
clusters of GPUs. 
\end{abstract}

\end{frontmatter}



\noindent \textbf{PROGRAM SUMMARY} 

\noindent \begin{small} {\em Manuscript Title:} ZMCintegral: a
Package for Multi-Dimensional Monte Carlo Integration on Multi-GPUs
\\
 {\em Authors:} Hong-Zhong Wu; Jun-Jie Zhang; Long-Gang Pang; Qun
Wang \\
 {\em Program Title: ZMCintegral} \\
 {\em Journal Reference:} \\
{\em Catalogue identifier:} \\
{\em Licensing provisions:} Apache License Version, 2.0(Apache-2.0)\\
{\em Programming language:} Python \\
{\em Operating system:} Linux 
{\em Classification:} 4.12 Other Numerical Methods \\
{\em External routines/libraries:} Tensorflow; Numba; Ray \\
{\em Nature of problem:} Easy to use python package for Multidimensional-multi-GPUs
Monte Carlo integration\\
{\em Solution method:} Stratified sampling and heuristic tree search
using multiple GPUs on distributed clusters\\

\section{Introduction}

Integrations of high dimensional functions are frequently encountered
in computional physics. For example, the total cross sections for
particle scatterings in high energy physics\cite{Peskin:1995ev,Zhang2019},
and the transport equations over phase space in many body physics\cite{DeGroot:1980dk,Xu2005,Chen2013},
etc. These integrations are usually time consuming due to the so called
curse of dimensionality\cite{Bellman1957,Bellman2003}; and often,
the ill-behaved integrands make the standard quadrature formulae infeasible.
Monte Carlo algorithms, with its non-deterministic intrinsic, is particularly
useful for higher-dimensional integrals.

The Monte Carlo integration method usually requires large sample points
to increase its calculation precision. One of the most popular Monte
Carlo algorithms for integration is VEGAS\cite{GPLepage1978}, where
the methods of importance sampling and adaptive stratified sampling
are applied and implemented on CPU with a user-friendly interface.
The newly versioned VEGAS, containing an adaptive multi-channel sampling
method (Ref. \cite{TOhl1999}), improves the accuracy for some typical
integrals, without introducing much longer evaluation time. However,
as the dimensionality increases, the required number of samples increases
exponentially to achieve sufficient precision. Therefore, a parallelization
of CPUs is needed to handle these large samples. It is proposed in
Ref. \cite{Kreckel1997}, that VEGAS, with a semi-micro-parallelization
method, can be utilized on multi-CPUs to increase speed. 

GPUs originally designed for accelerating high-quality computer video
games are proved to be very good at single instruction multiple data
parallelizations, where simple computation kernels are executed in
parallel on thousands of processing elements/threads that a single
GPU of a personal computer would have today. gVEGAS \cite{Kanzaki2011},
which parallelized VEGAS on GPU using CUDA \cite{cuda01,Nickolls2010},
brought $\approx50$ times performance boost compared to the CPU version.
In the same paper \cite{Kanzaki2011}, a program called BASES \cite{Kawabata1995}
which deals with the integration of singular functions is parallelly
performed on GPU as gBASES by KEK (The High Energy Accelerator Research
Organization in Japan). However, both gVEGAS and gBASES cannot currently
run in multi-GPU devices, and no easy-to-use API has been yet released. 

Foam \cite{Jadach2003}, another platform for Monte Carlo integration,
based on dividing the integration domain into small cells, is also
popular but lacks an official release of GPU supported version. Recently,
an improved method using Boosted Decision Trees and Generative Deep
Neural Networks suggested an advanced importance-sampling algorithm
for Monte Carlo integration \cite{Joshua2017}.

In this paper, we propose an easy-to-use python package, ZMCintegral,
for multi-dimensional Monte Carlo integration on distributed multi-GPU
devices. The source codes and manual can be found in Ref. \cite{zmc_github}.
It uses both stratified sampling and heuristic tree search algorithm
to perform Monte Carlo integration in each cell of the integration
domain. Speed and accuracy are both of our concern. It usually takes
about a few minutes to finish the evaluation and outputs an integral
result with an estimated standard deviation. The algorithm is scalable
as its speed is increased with the number of GPUs being used.

Currently, we have built three versions of ZMCintegral on multi-GPUs,
one based on Tensorflow eager mode \cite{Martin2015}\cite{Martin2016}\cite{tf_eager},
the other two based on Numba \cite{Kwan2015}. The Tensorflow version
wrapped the difficulties and complexities of multi-GPU parallelization,
such as correctly handle the CPU-GPU data transfer, the synchronization,
and dealing with the difference between global, local and private
memories of GPUs, deeply under beneath Google Tensorflow, which is
a python library developed originally for machine learning studies
and provides easy-to-use Python function interfaces. We have demonstrated
that this is a very good procedure for multi-GPUs parallelization
of scientific programs, especially for inexperienced scientific researchers.
The other two versions, based on Numba, are different in their parallelizing
methods. One uses python Multiprocessing library and focuses on one
node computation with multiple GPUs. The other uses Ray \cite{PMoritz2018}
and performs the calculation on multiple nodes. The numba versions
are highly optimized and flexible at a deep level, which makes high
dimensional integration feasible in a reasonable time. All the above
features make ZMCintegral an easy-to-use and high performance tool
for multi-dimensional integration. 

\section{The structure of ZMCintegeral}

\subsection{Stratified sampling and heuristic tree search}

In our calculation, the whole integration domain is divided into $k$
($k$ is an integer) equal sub-domains, with each sub-domain a hypercube
of the same volume. Then the integral in each hypercube is calculated
using the direct Monte Carlo method, and the integration process is
repeated independently by $n_{1}$ times to get $n_{1}$ independent
integration values in each sub-domain. For each integration in one
sub-domain, the total number of sampled points is $n_{2}^{D}$, where
$n_{2}$ is the number of points in each dimension for every sub-domain
and $D$ is the dimension of the integration. If the number of sampled
points is big enough, each of these $n_{1}$ integration values is
a good approximation of the real value for the specific sub-domain.
After $n_{1}$ independent integration value for each sub-domain is
calculated, the mean and standard deviation of these $n_{1}$ integration
values are computed for all $k$ sub-domains. For some of the sub-domains,
the standard deviation of $n_{1}$ integration is larger than the
threshold value (one hyper-parameter, and can be determined by the
mean and standard deviation of the calculated $k$ standard deviations),
indicating that the fluctuation is still too large and the integration
precision is not sufficient. As a result, these sub-domains need to
be recalculated by recursively being split into $k$ new smaller sub-domains.

For example, if the standard deviation in one sub-domain is larger
than the threshold value, the integrand in that sub-domain may strongly
fluctuate and the mean integration value may not be trusted, suggesting
that the integration has to be recalculated with higher accuracy.
For those sub-domains whose standard deviations are lower than the
threshold value, their mean integration values are accepted as the
real values of the sub-domain integration. For those sub-domains whose
integrations have to be recalculated, we divide each sub-domain further
into $k$ equal sub-sub-domains (depth 2). The same Monte Carlo calculation
is applied to these sub-sub-domains to obtain a new list of integration
values and corresponding standard deviations. Then a new threshold
value is set to filter out those sub-sub-domains that need to be recalculated.
This procedure continues until the standard deviations in all sub(-sub-$\cdots$)-domains
are lower than the threshold value or the maximal partition depth
is reached. Here the partition depth is defined as the total number
of layers the integration domain is divided. For example, if the partition
depth is 1, the integration domain would only be divided once and
sub-domains would not be further divided. Finally the accepted integration
values are collected in all sub-domains or sub(-sub-$\cdots$)-domains
to obtain the total integration value. At the same time, the standard
deviations are collected to obtain the integration error. The algorithm
is illustrated in Fig. \ref{fig:1}.
\begin{figure}[H]
\centering{}\includegraphics[scale=0.3]{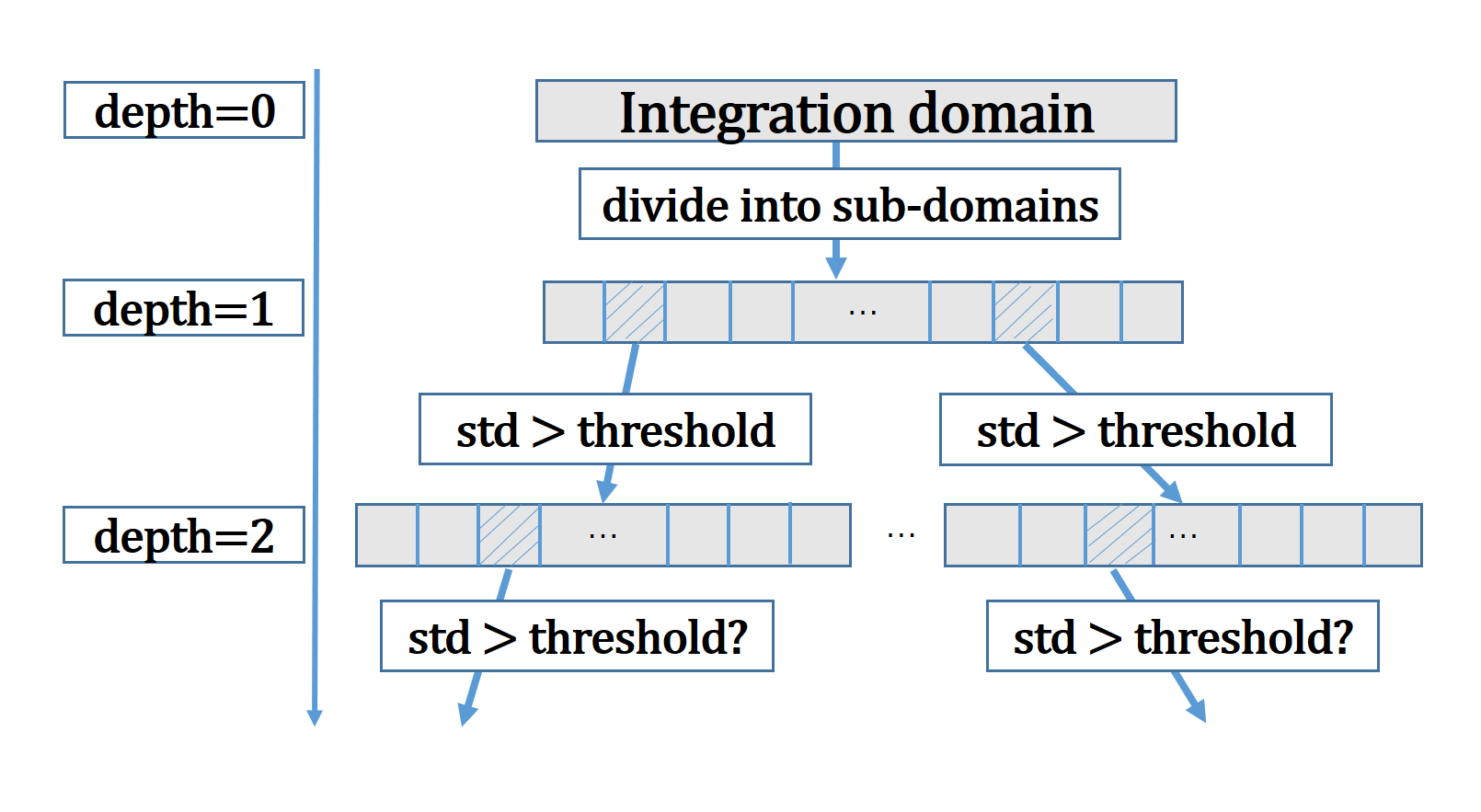}\caption{\label{fig:1}The illustration of the algorithm in ZMCintegral. The
integration domain is divided into sub-domains and a threshold value
is set to filter out those sub-domains whose integration values are
unstable and have to be recalculated. These unstable sub-domains are
further divided into sub-sub-domains and the integration in the sub-sub-domains
are calculated. The same procedure is continued until the integration
values in all sub(-sub-$\cdots$)-domains are accepted or the maximal
partition depth is reached.}
\end{figure}

\subsection{Usage of multi-GPUs}

One major difficulty to implement multi-dimensional integrations on
GPU with the stratified method is the limitation of graphic memory.
The reason is that the random numbers generated to calculate the integral
occupy a huge memory. The cost of graphic memory increases linearly
with the number of sub-domains. To solve this problem, we may apply
two methods: the first one is to use multi-GPUs to avoid insufficient
memory, where every GPU device only tackles limited sample points.
The second one is to modify the way we deal with random numbers generating\cite{Deak1990}:
instead of producing the random numbers in batch and storing them
in the graphic memory for later use, we only produce the random number
at the time when using it. For example, in calculating the integral
in one sub-domain, we produce the random number and calculate the
integral, and then iterate this produce-at-calculation process many
times by the number of sample points in the sub-domain. The advantage
of the produce-at-calculation method is that it allows a very large
number of sub-domains which improves the precision of the integration. 

We have three implementations of ZMCintegeral. One uses the functionality
of TensorFlow (TensorFlow version) on multi-GPU devices but without
improving the treatment of random numbers generating. Hence, when
the dimension of the integral is not very high (between 4 and 10)
the speed of the integration on multi-GPUs increases significantly;
however, when the dimension of the integral is very high (larger than
12), the GPU memory will not be enough. In this case, the method of
producing random numbers in real-time becomes more efficient. The
other two versions of ZMCintegral, a parallel computing package in
python on multi-GPUs, are realized with Numba. The difference between
the two numba versions is that one realization with multiprocessing
package (Numba-Multiprocessing version) can only be used on one node,
and the other with Ray \cite{PMoritz2018} (Numba-Ray version) can
be used in distributed clusters. The produce-at-calculation process
makes sampling a huge number of points possible so that the integration
precision is guaranteed. The speed of two the Numba versions are as
fast as the TensorFlow version.

\subsection{Parameters}

For different integrands with different dimensions, we provide some
parameters that can be adjusted by users. The typical hyper-parameters
are the number of independent repetitive evaluations for one sub-domain,
the threshold value above which the sub-domain has to be recalculated,
the maximal depth for tree search, and the number of GPUs that are
used for the calculation. Other parameters include the number of sub-domains
and the number of sample points in one sub-domain. The product of
these two parameters is limited by the computing resources. In the
TensorFlow version, it is preferred to increase the number of sample
points in one sub-domain, while in the Numba versions, it is preferred
to increase the number of sub-domains. The detailed parameters and
illustrations can be found in Ref. \cite{zmc_github}.

\section{Results and performance}

The performance of ZMCintegral is compared with VEGAS for two typical
functions: one is an oscillating function with 6 variables

\begin{equation}
f_{1}=\sin(\sum_{i=1}^{6}x_{i}),
\end{equation}
the other is a Gaussian function of 9 variables
\begin{equation}
f_{2}=\frac{1}{\left(\sqrt{2\pi}\sigma\right)^{9}}\exp\left(-\frac{1}{2\sigma^{2}}\sum_{i=1}^{9}x_{i}^{2}\right),
\end{equation}
where $\sigma=0.01$ and a high peak is located at $x_{i}=0$ for
$i=1,\cdots,9$. The integration domains of $f_{1}$ and $f_{2}$
are chosen to be $\{x_{i}\in[0,10],i=1,\cdots,6\}$ and $\{x_{i}\in[-1,1],i=1,\cdots,9\}$
respectively. As an illustration of the oscillation behavior of $f_{1}$
and the high peak feature of $f_{2}$, we plot the two functions with
two variables in Fig. \ref{fig:2}. 
\begin{figure}[H]
\centering{}\includegraphics[scale=0.3]{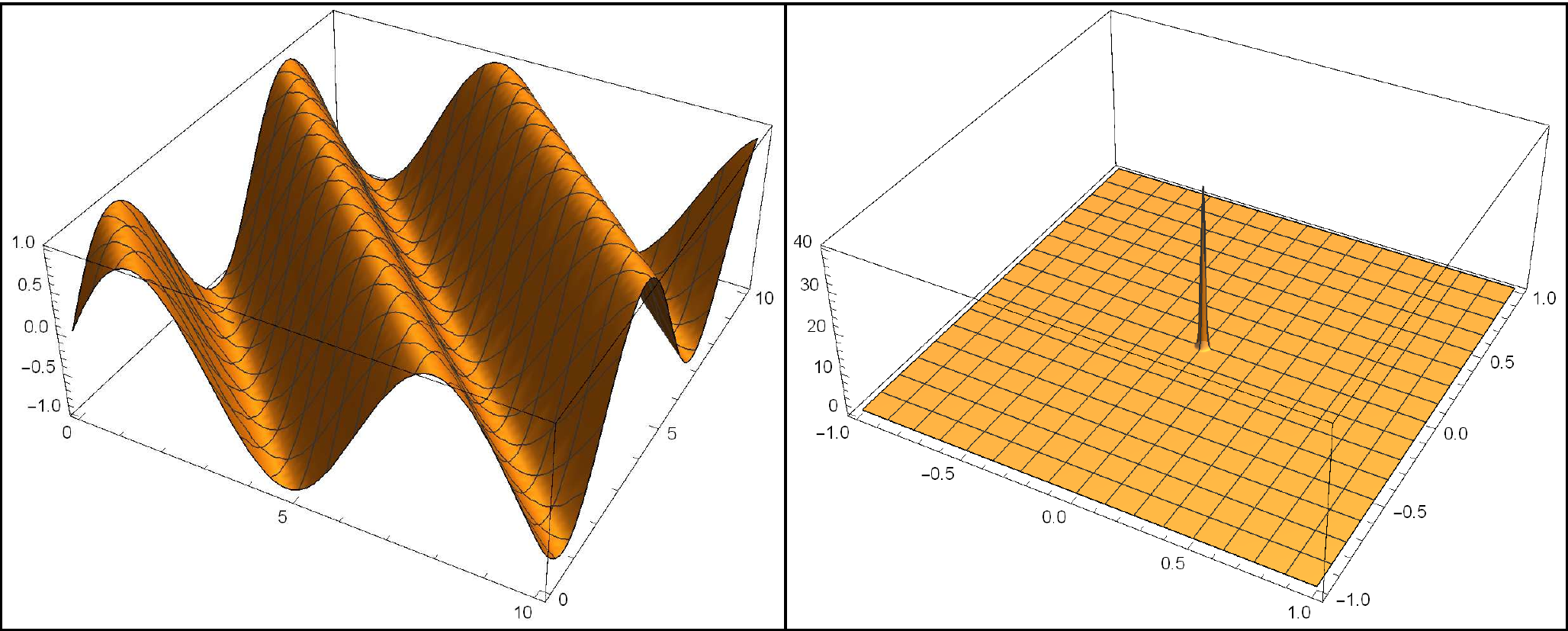}\caption{\label{fig:2}Schematic plot of sine and Gaussian functions with two
variables.}
\end{figure}

In our numerical experiments, the test platform is kept the same for
ZMCintegral and VEGAS. For TensorFlow and Numba-Multiprocessing versions,
we test them on one machine that can be treated as a single node.
The hardware condition for this node is Intel(R) Xeon(R) CPU E5-2620
v3@2.40GHz CPU with 24 processors + 4 Nvidia Tesla K40m GPUs. What
have to be mentioned here is that we only test the VEGAS on this node.
For the Numba-Ray version, we have tested it on multiple nodes. The
hardware condition for the three nodes are Intel(R) Xeon(R) CPU E5-2620
v3@2.40GHz CPU with 24 processors + 4 Nvidia Tesla K40m GPUs, Intel(R)
Xeon(R) CPU E5-2680 V4@2.40GHz CPU with 10 processors + 2 Nvidia Tesla
K80 GPUs, and Intel(R) Xeon(R) Silver 4110 CPU@2.10GHz CPU with 10
processors + 1 Nvidia Tesla V100 GPU. The K80 card can be seen as
the combination of two K40 cards in physical structure. These three
nodes are in a local area network.

\subsection{Test on one node}

The TensorFlow and Numba-Multiprocessing versions are tested in this
section. The test is performed on one node with 4 Nvidia Tesla K40m
GPU devices. The calculation results and the total evaluation time
is compared with that of VEGAS.

For the integral of $f_{1}$, we use both the TensorFlow and Numba-Multiprocessing
versions to carry out the integration and compare the results with
VEGAS. The parameters for TensorFlow version are set to the following
values: the number of sub-domains in one dimension is $3$, the sample
points in one sub-domain in one dimension is $20$, the number of
independent repetitive evaluations for one sub-domain is $5$, the
maximal depth is $2$, and sub-domains with standard deviations larger
than $5\sigma$ will be recalculated. We have sampled totally $20^{6}\times3^{6}\approx4.67\times10^{10}$
points for the 6-dimensional integration. For Numba-Multiprocessing
version, the parameters are set to the following values: the number
of sub-domains in one dimension is $6$, the sample points in one
sub-domain in one dimension is $10$, the number of independent repetitive
evaluations for one sub-domain is $5$, the maximal depth is $2$,
and sub-domains with standard deviations larger than $5\sigma$ will
be recalculated. Threads per block is chosen to be 16 initially, block
per grids is calculated via threads per block. The number of sample
points is same as the TensorFlow version but with more sub-domains
and less sample points in one sub-domain. In VEGAS, the calculation
is done on the same machine. We found that in order to obtain the
accepted precision, the number of sample points must at least be $10^{9}$,
as can be seen in Tab. \ref{tab:table-1} and Fig. \ref{fig:3}. We
use three modes in the calculation for VEGAS. One with iteration number
set to be $10$ and without discarding operation (normal VEGAS usage),
the second with the operation of discarding estimates where the initial
several iteration steps are discarded, the last one with both discarding
operation and the batch mode \cite{vegas_batch}.

To compare the stability, the total evaluation time and accuracy,
we list the averaged results of $10$ independent evaluations in Tab.
\ref{tab:table-1} and in Fig. \ref{fig:3}.
\begin{center}
\begin{table}[H]
\centering{}\caption{\label{tab:table-1}Results for the integral of $f_{1}$. Each data
is the averaged value of $10$ independent evaluations. In the most
left column, different versions of the packages are used. ZMC\_TF\_1K40m
means the tensorflow version with 1 K40m being used; and ZMC\_numba\_2K40m
means the Numba-Multiprocessing version with 2 K40m being used. VEGAS\_pre\_estimate
means VEGAS with discarding operation; and VEGAS\_batch means the
VEGAS discarding operation and the batch mode are being used.}
{\scriptsize{}}%
\begin{tabular}{ccccc}
\hline 
 & {\scriptsize{}calculation result} & {\scriptsize{}standard deviation} & {\scriptsize{}sample points} & {\scriptsize{}total time (s)}\tabularnewline
\hline 
\hline 
{\scriptsize{}ZMC\_TF\_1K40m} & {\scriptsize{}-48.96} & {\scriptsize{}0.76} & {\scriptsize{}$4.67\times10^{10}$} & {\scriptsize{}355.8}\tabularnewline
{\scriptsize{}ZMC\_TF\_2K40m} & {\scriptsize{}-49.20} & {\scriptsize{}1.61} & {\scriptsize{}$4.67\times10^{10}$} & {\scriptsize{}187.2}\tabularnewline
{\scriptsize{}ZMC\_TF\_3K40m} & {\scriptsize{}-49.02} & {\scriptsize{}1.36} & {\scriptsize{}$4.67\times10^{10}$} & {\scriptsize{}125.9}\tabularnewline
{\scriptsize{}ZMC\_TF\_4K40m} & {\scriptsize{}-49.05} & {\scriptsize{}1.13} & {\scriptsize{}$4.67\times10^{10}$} & {\scriptsize{}98.8}\tabularnewline
{\scriptsize{}ZMC\_numba\_1K40m} & {\scriptsize{}-49.56} & {\scriptsize{}1.40} & {\scriptsize{}$4.67\times10^{10}$} & {\scriptsize{}313.5}\tabularnewline
{\scriptsize{}ZMC\_numba\_2K40m} & {\scriptsize{}-49.20} & {\scriptsize{}1.15} & {\scriptsize{}$4.67\times10^{10}$} & {\scriptsize{}169.4}\tabularnewline
{\scriptsize{}ZMC\_numba\_3K40m} & {\scriptsize{}-49.58} & {\scriptsize{}1.03} & {\scriptsize{}$4.67\times10^{10}$} & {\scriptsize{}124.8}\tabularnewline
{\scriptsize{}ZMC\_numba\_4K40m} & {\scriptsize{}-49.09} & {\scriptsize{}0.61} & {\scriptsize{}$4.67\times10^{10}$} & {\scriptsize{}96.2}\tabularnewline
{\scriptsize{}VEGAS} & {\scriptsize{}-50.47} & {\scriptsize{}2.81} & {\scriptsize{}$10^{9}$} & {\scriptsize{}10238.5}\tabularnewline
{\scriptsize{}VEGAS\_pre\_estimate} & {\scriptsize{}-49.96} & {\scriptsize{}2.25} & {\scriptsize{}$10^{9}$} & {\scriptsize{}15647.3}\tabularnewline
{\scriptsize{}VEGAS\_batch} & {\scriptsize{}-48.94} & {\scriptsize{}1.98} & {\scriptsize{}$10^{9}$} & {\scriptsize{}3857.2}\tabularnewline
\hline 
\end{tabular}
\end{table}
\begin{figure}[H]
\centering{}\includegraphics[scale=0.5]{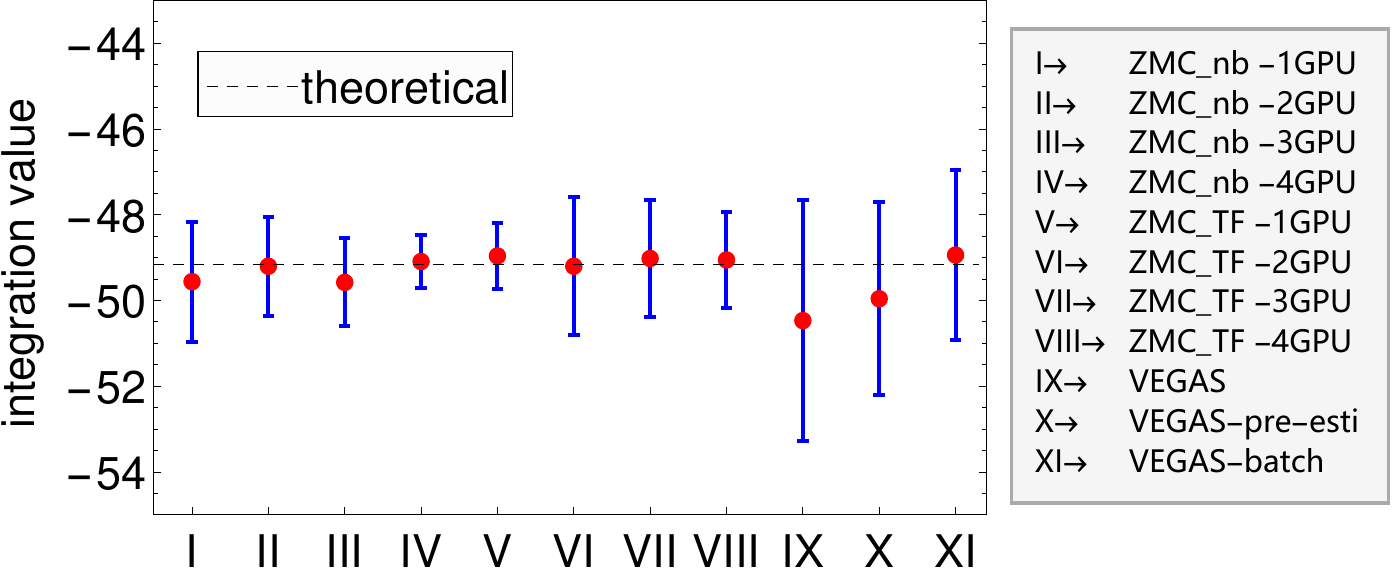}\caption{\label{fig:3}Results for the integral of $f_{1}$. Each data point
is the averaged value of $10$ independent evaluations. The data point
from left to right: results from the Numba-Multiprocessing version
with $4.67\times10^{6}$ sample points on 1 GPU, 2 GPUs, 3 GPUs, and
4 GPUs, results from the TensorFlow version on 1 GPU, 2 GPUs, 3 GPUs,
and 4 GPUs, the results from VEGAS with $10^{9}$ sample points without
discarding operation, with $5$ discarding operation, and with $5$
discarding operation and batch mode.}
\end{figure}
\par\end{center}

For the integral of $f_{2}$, we only compare the performances of
VEGAS and the Numba-Multiprocessing version (the TensorFlow version
requires more GPU memory than we can provide). The parameters are:
the number of sub-domains in one dimension is $3$, the sample points
in one sub-domain in one dimension is $3$, the number of independent
repetitive evaluations for one sub-domain is $5$, the maximal depth
is $4$, and sub-domains with standard deviations larger than $5\sigma$
will be recalculated. The number of total sample points is $3^{9}\cdot3^{9}\approx3.87\times10^{8}$.
In VEGAS, the number of integrand evaluation per iteration is $10^{7}$
such that the evaluation of the integral can be in the lowest cost.
The results are shown in \ref{tab:table-2} and Fig. \ref{fig:4}. 
\begin{center}
\begin{table}[H]
\centering{}\caption{\label{tab:table-2}Results for the integral of $f_{2}$.}
{\scriptsize{}}%
\begin{tabular}{ccccc}
\hline 
 & {\scriptsize{}calculation result} & {\scriptsize{}standard deviation} & {\scriptsize{}sample points} & {\scriptsize{}total time (s)}\tabularnewline
\hline 
\hline 
{\scriptsize{}ZMC\_numba\_1K40m} & {\scriptsize{}1.00001} & {\scriptsize{}0.00051} & {\scriptsize{}$3.87\times10^{8}$} & {\scriptsize{}69.7}\tabularnewline
{\scriptsize{}ZMC\_numba\_2K40m} & {\scriptsize{}0.99993} & {\scriptsize{}0.00048} & {\scriptsize{}$3.87\times10^{8}$} & {\scriptsize{}52.8}\tabularnewline
{\scriptsize{}ZMC\_numba\_3K40m} & {\scriptsize{}1.00016} & {\scriptsize{}0.00036} & {\scriptsize{}$3.87\times10^{8}$} & {\scriptsize{}45.1}\tabularnewline
{\scriptsize{}ZMC\_numba\_4K40m} & {\scriptsize{}1.00026} & {\scriptsize{}0.00045} & {\scriptsize{}$3.87\times10^{8}$} & {\scriptsize{}33.2}\tabularnewline
{\scriptsize{}VEGAS} & {\scriptsize{}0.99987} & {\scriptsize{}0.00050} & {\scriptsize{}$10^{7}$} & {\scriptsize{}510.0}\tabularnewline
\hline 
\end{tabular}
\end{table}
\begin{figure}[H]
\centering{}\includegraphics[scale=0.5]{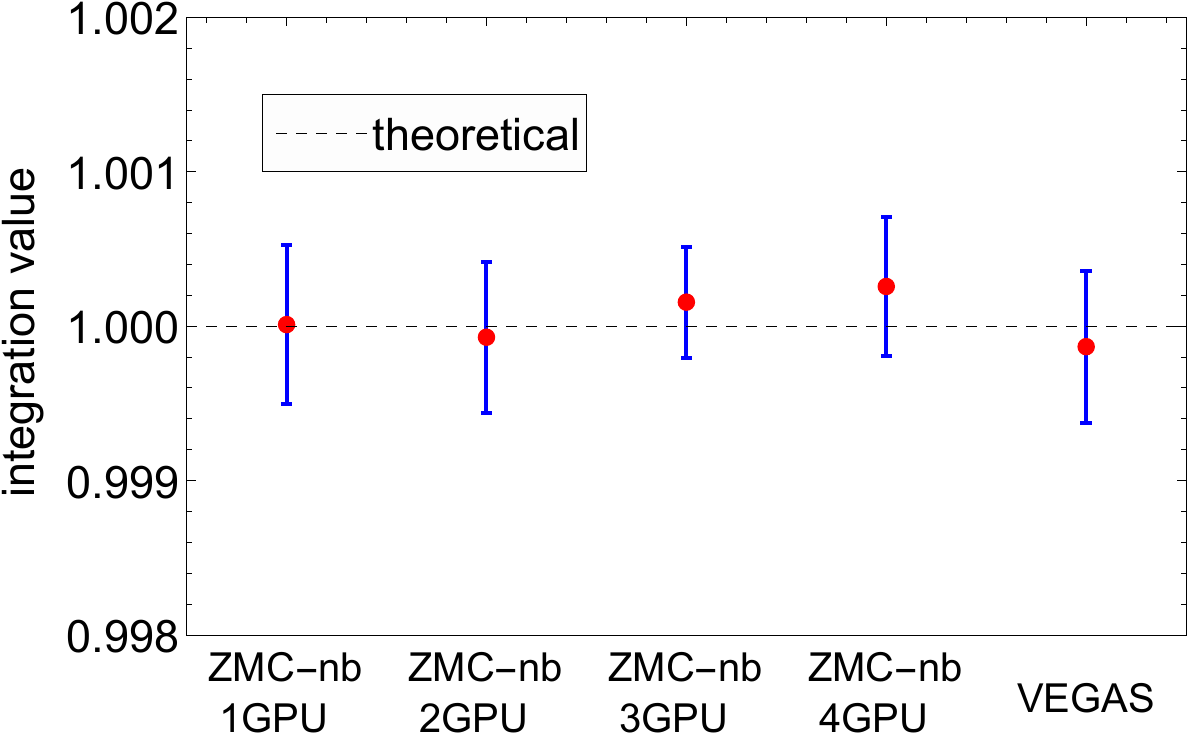}\caption{\label{fig:4}Results for the integral of $f_{2}$. Each data point
is the averaged value of $10$ independent evaluations. The data point
from left to right: results from the Numba-Multiprocessing version
of ZMCintegral on 1 GPU, 2 GPUs, 3 GPUs, and 4 GPUs, the result from
VEGAS with 5 steps of discarding operation.}
\end{figure}
\par\end{center}

\subsection{Test on multiple nodes}

The Numba-Ray version supports scalable computations on distributed
clusters with Ray \cite{PMoritz2018}. We give the detailed test for
this version on one node and multiple nodes. The integrands are choosen
to be $f_{1}$ and $f_{2}$ as well.

The parameters for $f_{1}$ is as follows: the number of sub-domains
in one dimension is $12$, the sample points in one sub-domain is
$10^{4}$, the number of independent repetitive evaluations for one
sub-domain is $5$, the maximal depth is $1$. The number of sample
points is $12^{6}\times10^{4}\approx2.99\times10^{10}$. Threads per
block is choosen to be 32, and we have found very little difference
with other values. In our experiments, we have recorded the averaged
calculation results, the standard deviation and the total evaluation
time. Besides this basic information, we also keep track of the time
consumption for tasks allocating and data retrieving between the head
node and remote nodes through the network, as well as the evaluation
time for single GPU card for one call. The data transfer time between
host and device is also recorded. The results are shown in Tab. \ref{tab:table-3}
and in Fig. \ref{fig:5},\ref{fig:6},\ref{fig:7},\ref{fig:8}.
\begin{center}
\begin{table}[H]
\centering{}\caption{\label{tab:table-3}Integration results for $f_{1}$ with different
node configurations. 4 K40m means the calculation is done on one node
which uses four K40m. 4 K40m + 2 K80 means the calculation is done
using two nodes, one node uses four K40m and the other uses two K80.
For evaluations using more than two nodes, the time for Host to Device
and Device to Host are monitored seperately for different GPU. For
example, 10.7/19.0/20.2 means 10.7 $\text{\ensuremath{\mu}s}$ for
one K40m per call, 19.0 $\text{\ensuremath{\mu}s}$ for one K80 per
call and 20.2 $\text{\ensuremath{\mu}s}$ for one V100 per call. GPU
Calc means the evaluation time per GPU for one call.}
{\scriptsize{}}%
\begin{tabular}{ccccc}
\hline 
 & {\scriptsize{}calculation result} & {\scriptsize{}standard deviation} & {\scriptsize{}total time(s)} & {\scriptsize{}allocate(ms)}\tabularnewline
\hline 
\hline 
{\scriptsize{}1K40m} & {\scriptsize{}-49.4674} & {\scriptsize{}0.8606} & {\scriptsize{}298.8} & {\scriptsize{}6.5}\tabularnewline
{\scriptsize{}2K40m} & {\scriptsize{}-49.3327} & {\scriptsize{}0.4885} & {\scriptsize{}122.5} & {\scriptsize{}6.7}\tabularnewline
{\scriptsize{}3K40m} & {\scriptsize{}-49.3619} & {\scriptsize{}0.5379} & {\scriptsize{}57.0} & {\scriptsize{}5.9}\tabularnewline
{\scriptsize{}4K40m} & {\scriptsize{}-49.4126} & {\scriptsize{}0.7318} & {\scriptsize{}35.1} & {\scriptsize{}5.9}\tabularnewline
{\scriptsize{}1V100} & {\scriptsize{}-49.3971} & {\scriptsize{}0.4845} & {\scriptsize{}38.7} & {\scriptsize{}3.8}\tabularnewline
{\scriptsize{}2K80} & {\scriptsize{}-49.2317} & {\scriptsize{}0.6695} & {\scriptsize{}38.9} & {\scriptsize{}6.3}\tabularnewline
{\scriptsize{}4K40m+2K80} & {\scriptsize{}-49.4891} & {\scriptsize{}0.5896} & {\scriptsize{}20.2} & {\scriptsize{}7.5}\tabularnewline
{\scriptsize{}4K40m+2K80+1V100} & {\scriptsize{}-49.5273} & {\scriptsize{}0.5161} & {\scriptsize{}18.4} & {\scriptsize{}8.0}\tabularnewline
\hline 
 & {\scriptsize{}retrieve(ms)} & {\scriptsize{}HtoD($\text{\ensuremath{\mu}s}$)} & {\scriptsize{}GPU Calc(s)} & {\scriptsize{}DtoH(ms)}\tabularnewline
\hline 
\hline 
{\scriptsize{}1K40m} & {\scriptsize{}47.0} & {\scriptsize{}12.6} & {\scriptsize{}2.24} & {\scriptsize{}0.87}\tabularnewline
{\scriptsize{}2K40m} & {\scriptsize{}48.9} & {\scriptsize{}12.4} & {\scriptsize{}2.17} & {\scriptsize{}0.81}\tabularnewline
{\scriptsize{}3K40m} & {\scriptsize{}42.6} & {\scriptsize{}13.4} & {\scriptsize{}2.03} & {\scriptsize{}0.90}\tabularnewline
{\scriptsize{}4K40m} & {\scriptsize{}48.8} & {\scriptsize{}10.1} & {\scriptsize{}1.69} & {\scriptsize{}0.62}\tabularnewline
{\scriptsize{}1V100} & {\scriptsize{}45.7} & {\scriptsize{}16.3} & {\scriptsize{}0.46} & {\scriptsize{}0.76}\tabularnewline
{\scriptsize{}2K80} & {\scriptsize{}59.5} & {\scriptsize{}15.9} & {\scriptsize{}1.83} & {\scriptsize{}0.83}\tabularnewline
{\scriptsize{}4K40m+2K80} & {\scriptsize{}761.4} & {\scriptsize{}10.5/13.1} & {\scriptsize{}1.71/1.82} & {\scriptsize{}0.68/0.61}\tabularnewline
{\scriptsize{}4K40m+2K80+1V100} & {\scriptsize{}760.7} & {\scriptsize{}10.7/19.0/20.2} & {\scriptsize{}1.70/1.86/0.45} & {\scriptsize{}0.73/0.68/0.77}\tabularnewline
\hline 
\end{tabular}
\end{table}
\begin{figure}[H]
\centering{}\includegraphics[scale=0.5]{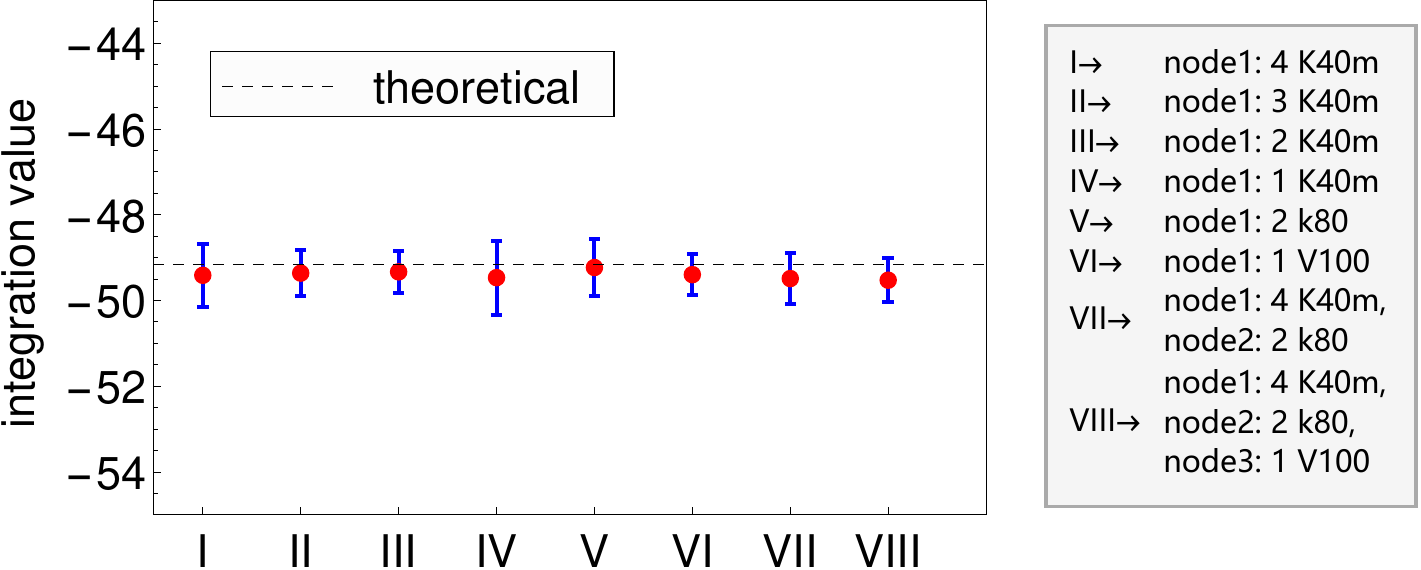}\caption{\label{fig:5}Integration results for $f_{1}$ with different node
configurations. Each data point is the averaged value of $10$ independent
evaluations. The data point from left to right corresponds to: one
node with 4 K40m, one node with 3 K40m, one node with 2 K40m, one
node with 1 K40m, one node with 2 K80, one node with 1 V100, two nodes
with 4 K40m + 2 K80, three nodes with 4 K40m +2 K80 +1 V100.}
\end{figure}
\par\end{center}

\begin{center}
\begin{figure}[H]
\centering{}\includegraphics[scale=0.4]{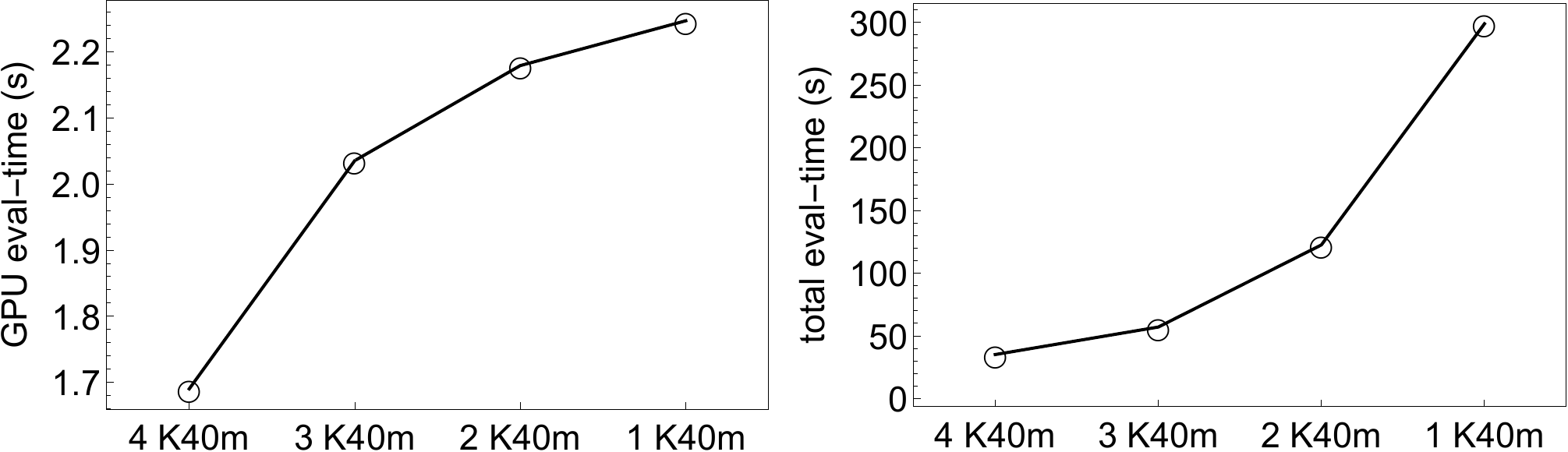}\caption{\label{fig:6} Time consumption for different number of GPUs on one
node. The left panel shows that when less GPUs are used, the evaluation
time per GPU for one call is increasing. This small time deivation
will add up to a significant time difference for the total evaluation
time, as can be seen in the right panel. The right panel shows that
when less GPUs are used, the total evaluation time is not linearly
increasing. This non-linearity suggests that more GPUs should be used
on one node.}
\end{figure}
\begin{figure}[H]
\centering{}\includegraphics[scale=0.4]{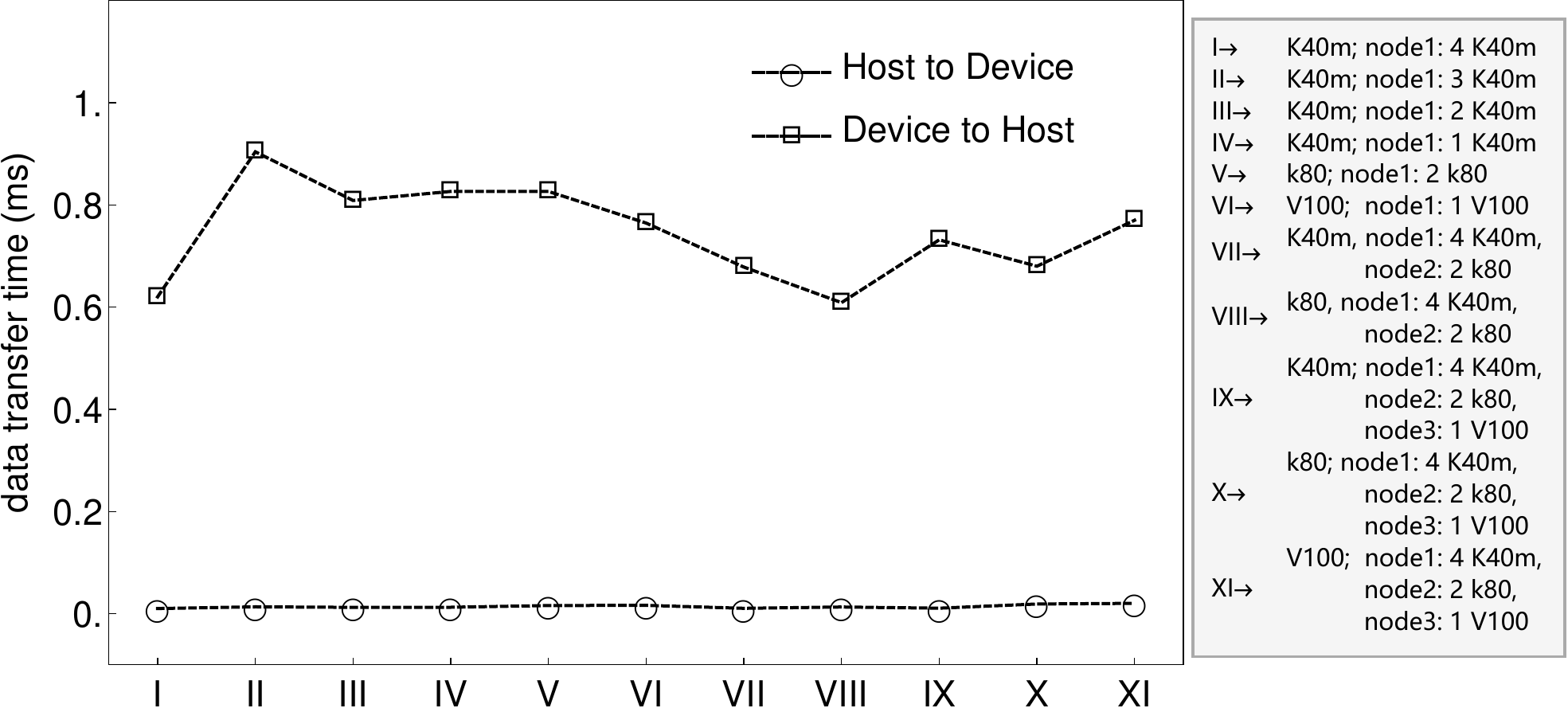}\caption{\label{fig:7}Data transfer time between GPU (device) and CPU (host)
with different node configurations. The data point from left to right
corresponds to: monitor of each K40m per call when one node with 4
K40m, 3 K40m, 2 K40m or 1 K40m is used, monitor of each K80 per call
when one node with 2 K80 is used, monitor of each V100 per call when
one node with 1 V100 is used, monitor of each K40m and K80 per call
when two nodes with 4 K40m + 2 K80 are used, monitor of each K40m,
K80 and V100 per call when three nodes with 4 K40m +2 K80 +1 V100
are used.}
\end{figure}
\par\end{center}

\begin{center}
\begin{figure}[H]
\centering{}\includegraphics[scale=0.3]{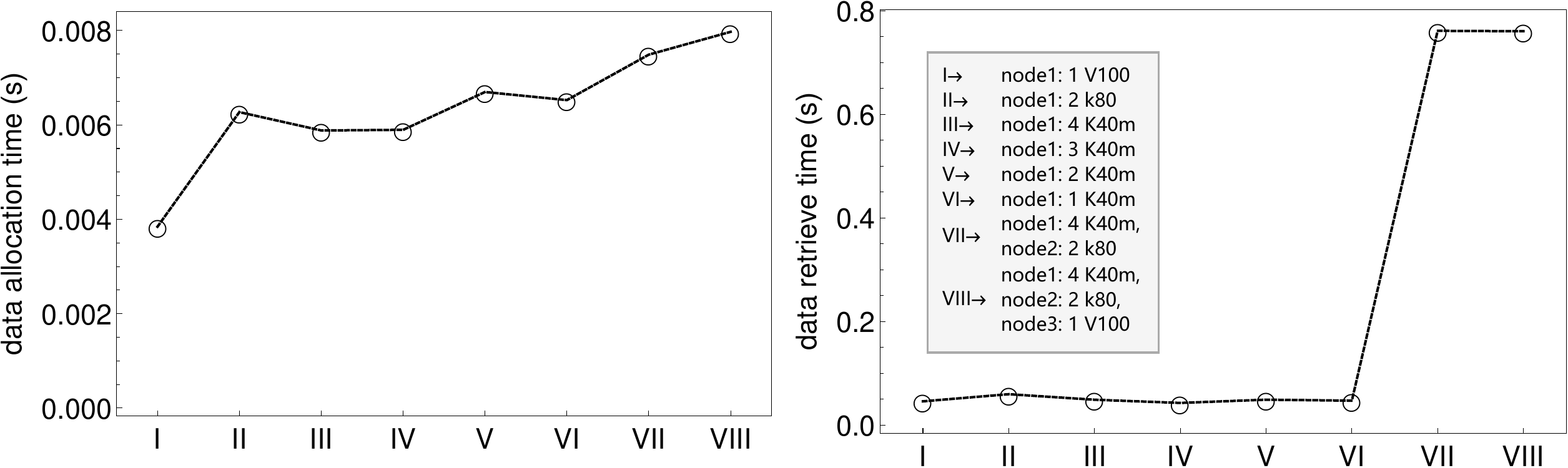}\caption{\label{fig:8}Data allocation and retrieve time from head node to
remote nodes with different node configurations. The left panel shows
the data allocation time from head to remote nodes via internet connection.
It can be seen that more time is needed when more nodes are being
involved. The right panel shows the data retrieve time from remote
nodes to head node. The ``jump'' for the last two configurations
suggests that when more nodes are used the retrieve time is increased.}
\end{figure}
\par\end{center}

For the Gaussian type integration, we test the performance for 9 dimensional
integration of $f_{2}$. The parameters are: the number of sub-domains
in one dimension is $3$, the sample points in one sub-domain is $10^{4}$,
the number of independent repetitive evaluations for one sub-domain
is $5$, the maximal depth is $4$, and sub-domains with standard
deviations larger than $4\sigma$ will be recalculated. The number
of sample points is $3^{9}\times10^{4}\approx1.97\times10^{8}$. The
test for the Gaussian type of 12 dimensions is also performed. The
results are in Tab. \ref{tab:table-4} and Fig. \ref{fig:9}. 
\begin{table}[H]
\centering{}\caption{\label{tab:table-4} Performance on 9-D and 12-D Gaussian integrals.
4 K40m\_9D means one node with four K40m are used for 9-D integration
and similar for the rest.}
\begin{tabular}{ccccc}
\hline 
 & {\scriptsize{}calculation result} & {\scriptsize{}standard deviation} & {\scriptsize{}total time(s)} & {\scriptsize{}allocate(ms)}\tabularnewline
\hline 
\hline 
{\scriptsize{}4K40m\_9D} & {\scriptsize{}0.99989} & {\scriptsize{}0.00059} & {\scriptsize{}8.2} & {\scriptsize{}0.8}\tabularnewline
{\scriptsize{}2K80\_9D} & {\scriptsize{}1.00015} & {\scriptsize{}0.00065} & {\scriptsize{}9.0} & {\scriptsize{}1.2}\tabularnewline
{\scriptsize{}1V100\_9D} & {\scriptsize{}1.00035} & {\scriptsize{}0.00064} & {\scriptsize{}4.8} & {\scriptsize{}0.9}\tabularnewline
{\scriptsize{}1V100\_12D} & {\scriptsize{}1.00002} & {\scriptsize{}0.00079} & {\scriptsize{}47.6} & {\scriptsize{}1.7}\tabularnewline
\hline 
 & {\scriptsize{}retrieve(ms)} & {\scriptsize{}HtoD(us)} & {\scriptsize{}GPU Calc(s)} & {\scriptsize{}DtoH(ms)}\tabularnewline
\hline 
\hline 
{\scriptsize{}4K40m\_9D} & {\scriptsize{}3.9} & {\scriptsize{}10.0} & {\scriptsize{}0.36} & {\scriptsize{}0.50}\tabularnewline
{\scriptsize{}2K80\_9D} & {\scriptsize{}4.4} & {\scriptsize{}12.1} & {\scriptsize{}0.37} & {\scriptsize{}0.58}\tabularnewline
{\scriptsize{}1V100\_9D} & {\scriptsize{}3.8} & {\scriptsize{}15.8} & {\scriptsize{}0.16} & {\scriptsize{}0.66}\tabularnewline
{\scriptsize{}1V100\_12D} & {\scriptsize{}6.9} & {\scriptsize{}15.7} & {\scriptsize{}0.73} & {\scriptsize{}0.72}\tabularnewline
\hline 
\end{tabular}
\end{table}
\begin{figure}[H]
\centering{}\includegraphics[scale=0.5]{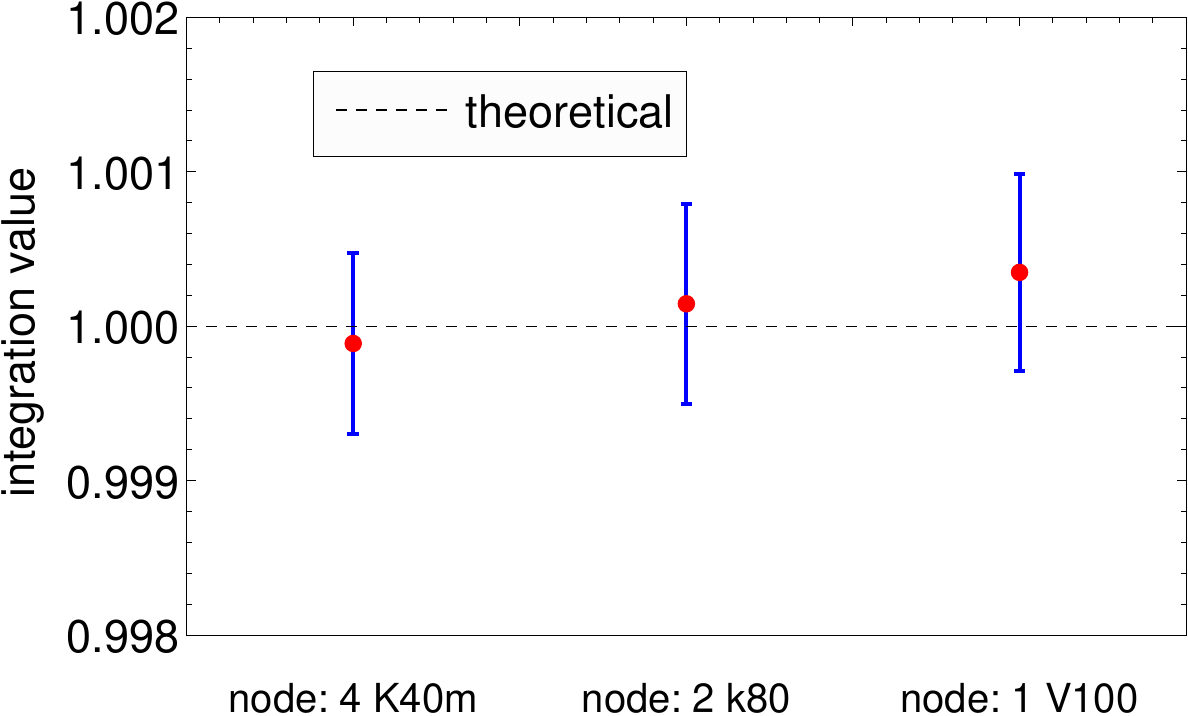}\caption{\label{fig:9}Results for the integral of $f_{2}$ with different
node configurations. Each data point is the average value of $10$
independent evaluations. The data point from left to right: 4 K40m
as one node, 2 K80 as one node, and V100 as one node.}
\end{figure}

\subsection{Analysis of the results}

Under a similar precision, ZMCintegral is much faster in speed than
VEGAS in our experiments. We can see in Tab. \ref{tab:table-1} that
for 6 dimensional integration, with the number of sample points being
$46$ times larger than that of VEGAS, ZMCintegral is still 10-150
times faster than VEGAS. For the 9 dimensional integration, as is
shown in Tab. \ref{tab:table-2}, when the number of sample points
is $38$ times larger than that of VEGAS, the speed of ZMCintegral
is still 7-15 times faster.

In the test of multiple nodes, we can see that the integration speed
is increased when more GPUs (more than 4) are used, as is shown in
Tab. \ref{tab:table-3}. From Fig. \ref{fig:6}, it is suggested that
even on one node more GPUs should be used. The data transfer time
between host to device is mainly dominated by the GPU cards. As can
be seen from Fig. \ref{fig:7}, for different configurations of nodes,
the time for HtoD (host to device) and DtoH (device to host) almost
kept unchanged.

While VEGAS is more efficient to put sample points in integration
domains where the integrand fluctuates dramatically, the GPU backend
ZMCintegral with stratified sampling method can put sample points
in each sub-domain whose number is comparable to the number of points
been put into the emphasized regions of VEGAS. Meanwhile, the use
of the heuristic tree search algorithm assures more sample points
in sub-domains with more fluctuations. Furthermore, ZMCintegral has
an appealing feature: the computation is scalable and its speed is
increased with the increasing number of GPUs in usage. While VEGAS
often needs a discarding operation process to ensure the reliability
of the final result, ZMCintegral only needs an appropriate value of
the maximal depth (usually between 2 and 4). In ZMCintegral, we indeed
see the effect of the heuristic tree search which yields a more precise
result than without the tree search.

A straight forward application of the two Numba version (Numba-Multiprocessing
and Numba-Ray) of ZMCintegral is the calculation of the global polarization
in heavy ion collisions. There we encounter an oscillating 10-dimensional
integration involving three momenta of the two incoming and outgoing
particles, where the incoming particles are wave-packets centered
at $\mathbf{p}_{A}$ and $\mathbf{p}_{B}$ and the outgoing particles
are plane waves with momenta $\mathbf{p}_{1}$ and $\mathbf{p}_{2}$.
With proper conservation laws the integration really involves $p_{1,x},p_{1,y},p_{1,z},p_{2,x},p_{2,y},p_{2,z},k_{A,y},k_{A,z},k_{A,y}^{\prime}$
and $k_{A,z}^{\prime}$, where $\mathbf{k}_{A}$ and $\mathbf{k}_{A}^{\prime}$
are the quantum fluctuations about $\mathbf{p}_{A}$. The difficulty
for this massive (containing several thousands terms) 10-dimensional
integration mainly comes from two aspects. On the one hand, a sufficient
number of sample points are required to cover all domains so that
the oscillating details should not be smeared out. To be safe, we
need around $15^{10}\approx5.77\times10^{11}$ sample points, which
is almost impossible for normal CPU algorithm. On the other hand,
the complexity of the integrand requires a very flexible interface
that all the condition checking, pattern matchings and special functions
can be easily realized. This we have done with Numba language. The
two Numba version of ZMCintegral, with full support of the Numpy and
Math packages, is a convenient choice.

\section{Conclusion and Discussion}

We propose a multi-GPU backend package for Monte Carlo integration
using stratified sampling and heuristic tree search algorithm. The
package has good performance for integration in large dimensions.
As can be seen in our examples, the package is able to control both
the time consumption and the accuracy, and it is tens to hundreds
times as fast as the traditional CPU based VEGAS. ZMCintegral is really
of great help for high dimensional integrations in large scale scientific
computation in terms of shortening the period of computing time from
months to days.

Instead of manipulating GPU with CUDA directly in C++, we choose the
state-of-art multi-GPU parallelization libraries Tensorflow and Numba
to build ZMCintegral. Both Tensorflow and Numba are professional and
widely used in the community and are easy to use with Python for non-experts.

Though ZMCintegral is able to handle integrations in dimensions up
to 16 and even higher, we should also keep in mind its limitation
for very high dimensions, for example, above 20. The difficulty lies
in the sampling. It was reported by Pan in MIT China submit \cite{Pan2018},
that we are not even able to deal with $2^{80}$ numbers with all
computational resources in the world within a year. So a general Monte
Carlo integration based on large scale sampling for very high dimensions
has a ceiling, in this case we may need other ``renormalization''
algorithms to reduce the dimension.

\section*{\textit{Acknowledgment}.}

HZW, JJZ and QW are supported in part by the Major State Basic Research
Development Program (973 Program) in China under Grant No. 2015CB856902
and by the National Natural Science Foundation of China (NSFC) under
Grant No. 11535012. LGP is supported by NSF under grant No. ACI-1550228
within the JETSCAPE Collaboration. The Computations are performed
at the GPU servers of department of modern physics at USTC.

\end{small}


\label{}



\label{-1}





  \bibliographystyle{elsarticle-num}
\nocite{*}
\bibliography{ZMCintegral}






\end{document}